# Toward the Formation of Realistic Galaxy Disks

Alyson Brooks

*Theoretical Astrophysics, Caltech, MC 350-17, Pasadena, CA 91125*

**Abstract.**
In this review I demonstrate that a realistic model for the formation of galaxy disks depends on a proper treatment of the gas in galaxies. Historically, cosmological simulations of disk galaxy formation have suffered from a lack of resolution and a physically motivated feedback prescription. Recent computational progress has allowed for unprecedented resolution, which in turn allows for a more realistic treatment of feedback. These advances have led to a new examination of gas accretion, evolution, and loss in the formation of galaxy disks. Here I highlight the role that gas inflows, the regulation of gas by feedback, and gas outflows play in achieving simulated disk galaxies that better match observational results as a function of redshift.

## 1.    The Role of Gas in the Formation of Disk Galaxies

Stellar light provides us with most of our information on galaxy evolution over the history of the Universe. The ability to use this stellar light to trace the evolution of disk galaxies depends critically on our model of how gas gets into galaxies and when and where stars are subsequently allowed to form from that gas. Unfortunately, the evolution of gas within dark matter halos in our Cold Dark Matter (CDM) dominated Universe is complicated by our inability to accurately model complex gas processes that allow it to cool, to heat, and to be shocked, so that it is unlikely to follow the evolution of the dark matter. In the simplest analytic models for galaxy formation, the virialization of the dark matter halo shock heats the gas to the virial temperature of the resulting halo. Any subsequently accreted gas is then also shock heated to the virial temperature of the halo. After halo collapse, the baryons are able to cool and condense to the center of the dark matter halo after adopting some density profile (e.g., White & Frenk 1991; Kauffmann et al. 1993; Cole et al. 2000; Somerville et al. 2001). The angular momentum of the gas and dark matter is thought to be accumulated by tidal torques exerted by neighboring structures (e.g., White 1984; Barnes & Efstathiou 1987), (though see Maller et al. 2002; Vitvitska et al. 2002).

Because mergers, supernova (SN) feedback, and gas accretion can be self-consistently modeled in simulations, the results from simulations can be used to test predictions from analytic models and to fine tune the physical prescriptions that are used by modelers. However, before simulation results can be used to better model galaxy formation and evolution, simulations must be tested against existing scaling relations simultaneously to establish robust constraints. Historically, matching these relations has not been a trivial task. In particular, simulated disks were too compact with a large central mass concentration or large stellar spheroid (e.g., Steinmetz & Navarro 1999; Navarro & Steinmetz 2000; Eke et al. 2001; Abadi et al. 2003; Governato et al. 2004).

This review details a new way of thinking about the evolution of gas in galaxies and the subsequent growth of the stellar disk that has emerged in recent years due to results from simulations. However, these results first require that the models are capable of creating realistic disks, and I thus begin in §2 with a discussion of the role that numerical resolution and a physically motivated feedback model play in achieving realistic simulated disks.[1] I then summarize in §3 a few of the key things that we have learned about gas accretion and its subsequent evolution from state-of-the-art simulations. In particular, dense filaments of gas can penetrate inside the shock heated gas halo of massive galaxies at high z. This "cold flow" gas can cool faster to the disk and form stars at higher redshifts than predicted by the standard model. In §4 I emphasize that it is essential to regulate star formation (SF) from this accreted gas in order to reproduce observed galaxy trends such as metallicity and gas content as a function of redshift. Finally, in §5 I turn to gas loss, demonstrating that a realistic model for SF and feedback leads to gas outflows from galaxies that alter the angular momentum distribution of the baryons, and even flatten the initial dark matter density profile, reconciling the discrepancies between observations and theory. I conclude in §6 with prospects for future work that still need to be addressed.

## 2.    Achieving Realistic Disks in Cosmological Simulations

The Tully-Fisher relation is a clear trend between the luminosity and rotational velocity of observed disk galaxies (e.g., Giovanelli et al. 1997; Geha et al. 2006), and provides a test for any simulated disk galaxy to match. However, historically simulated disks have been too compact (dense), too small overall in radius, and rotate too quickly at a given luminosity, making them unable to lie on the observed Tully-Fisher relation (see for example Steinmetz & Navarro 1999; Navarro & Steinmetz 2000; Eke et al. 2001). This problem has been termed the "angular momentum catastrophe," and it is significantly aggravated by another related problem, the "overcooling" problem.

In the overcooling problem, baryons cool too quickly at early times and become very dense and concentrated at the center of halos. These galaxies with an early, dense concentration of baryons undergo subsequent mergers within CDM. Merging subhalos experience dynamical friction, and orbital angular momentum is transferred to the dark matter of the accreting halo. By the time the dense baryons arrive at the disk, they have little angular momentum left, and the resulting disks then show the classic signs of the angular momentum catastrophe (Navarro & Benz 1991; Navarro & White 1994; Katz et al. 1994; Maller & Dekel 2002).

---

[1] While this review is focused on N-Body + SPH models, which have dominated the field of cosmological disk simulations to z = 0, significant progress has been made recently in running Adaptive Mesh Refinement (AMR) models to z = 0 (e.g., Gibson et al. 2009), and results from the two models are converging (O'Shea et al. 2005).

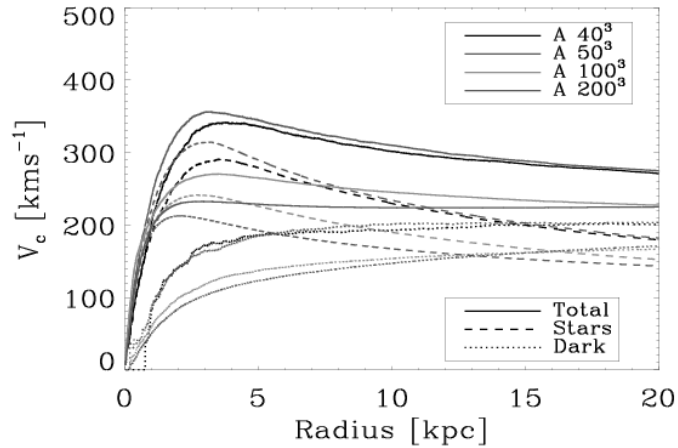

Figure 1.    Circular velocities versus radius, as a function of resolution, from Naab et al. (2007), reproduced by permission of the AAS. $V_c$ traces the mass interior to a given radius. It can be seen that as the number of particles is increased, $V_c$ flattens and the central mass concentration is reduced. These simulations were run without feedback, highlighting the role of resolution.

As discussed below, low numerical resolution can significantly contribute to the angular momentum catastrophe, but an inability to correctly model SF and feedback may be the larger contributor, via the overcooling problem.

## 2.1.    The Role of Resolution in Simulating Disks

In N-Body + Smoothed Particle Hydrodynamics (SPH) simulations, the dark matter (N-Body) particles are typically an order of magnitude more massive than the gas (SPH) particles. When a massive dark matter particle passes through a group of gas particles in the disk that have net rotation, there is a dramatic exchange of angular momentum in two body interactions.[2] This increases the random motions of the disk particles and relaxes the disk into a more spheroidal distribution. This loss of angular momentum also causes the gas to be more centrally concentrated, resulting in disks that are smaller and more compact as resolution is decreased (Kaufmann et al. 2007; Mayer et al. 2008). Simulated disks with sizes comparable to observed disk sizes do not form until a minimum of $10^6$ particles is reached (Governato et al. 2004; Kaufmann et al. 2007; Governato et al. 2007; Mayer et al. 2008).

Historically, it has been difficult to reach such resolutions in fully cosmological simulations, which require a much larger volume box and many more particles. To achieve higher resolution for a single halo, modelers must use what is known as the "volume renormalization" or the "zoom-in" technique (Katz

---

[2]This is only one source of artificial angular momentum loss due to numerical effects. See Mayer et al. (2008) for a full discussion. Also note that some angular momentum loss in real galaxies will occur due to torques when asymmetries are present or due to actual clumpy accretion, but the effect discussed here is artificial.

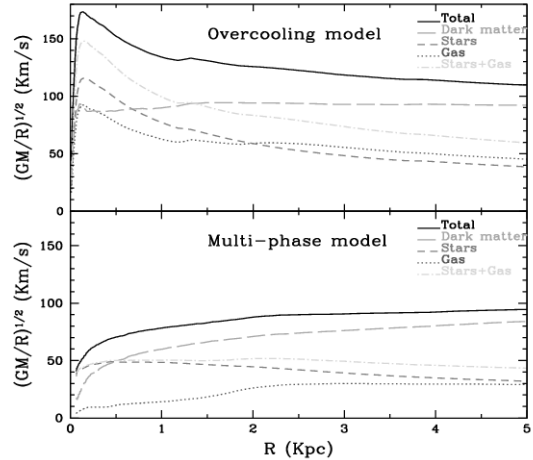

Figure 2.     Circular velocities versus radius, as a function of feedback prescription, from Ceverino & Klypin (2009), reproduced by permission of the AAS. The top panel shows that mass becomes too centrally concentrated if feedback is inefficient (and gas "overcools"). The bottom panel shows results for the same galaxy, but run with a SN feedback scheme that efficiently heats the gas and dramatically reduces the central concentration of baryons.

& White 1993; Navarro & White 1994). In this case, halos to be examined at higher resolution are selected from a large cosmological N-Body volume. Successively finer resolution layers of dark matter are added to the initial conditions around the selected galaxy, until at the finest layer gas particles are also added. This technique allows modelers to reach very high resolution in the simulated galaxy while retaining the torques due to large scale structure, but at a much cheaper computational cost than if we were to run the entire box at the highest resolution. Despite this, only a handful of cosmological simulations in the literature have achieved more than a million particles within the virial radius of a simulated galaxy.

Fig. 1, from Naab et al. (2007), shows the effect of resolution on the circular velocity of a galaxy as a function of radius, i.e., the amount of mass interior to that radius. These simulations were run without feedback, and so highlight the effect of resolution. The amount of mass concentrated in the center of the galaxy is dramatically reduced as resolution is increased (see also Governato et al. 2008; Piontek & Steinmetz 2009), e.g., $V_c$ drops ~100 km/s at 5 kpc as the number of particles is increased from $50^3$ to $200^3$, reducing the amount of mass interior by an order of magnitude. However, as suggested by Fig. 1, the central portions of simulated galaxies remain a challenge, with $V_c$ failing to converge in this region even at the highest resolutions. Despite high resolutions, the simulated bulges of simulated galaxies are still much too large compared to the bulges of real disk galaxies. It will be shown below that the SF and feedback prescription has a much more significant effect on these central regions.

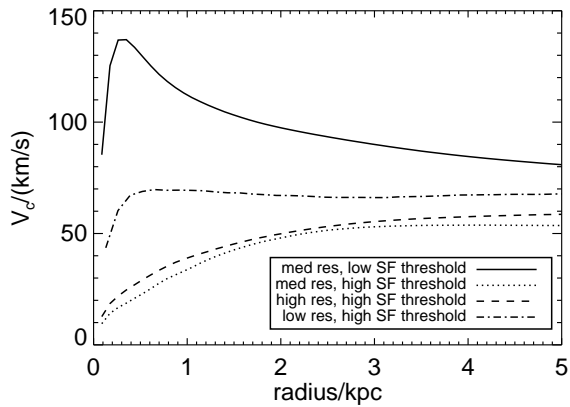

Figure 3.    Comparison of the circular velocity versus radius at $z=0$ for the low density SF threshold, and the high density star formation threshold at low, medium (3x higher force resolution and 27x higher mass resolution), and high (4x higher force and 64x higher mass) resolutions for the same low mass disk galaxy. The high density SF run leads to gas outflows that significantly reduce the central mass concentration. The SF scheme has much more significant effect on the mass distribution than resolution.

## 2.2.    The Role of Feedback

To prevent the overcooling problem and its contribution to the angular momentum catastrophe, feedback mechanisms have traditionally been invoked (White & Rees 1978; Dekel & Silk 1986; Navarro & Benz 1991; Navarro & White 1994; Katz et al. 1994; Maller & Dekel 2002). Feedback serves to prevent the rapid, early cooling of gas in simulated halos. This heating also "puffs up" the gas, moving it to a larger radius and increasing its angular momentum. The creation of a hot gas reservoir prevents early cooling, so that gas cools onto galaxies at later times, after the period of rapid mergers. Thus the baryons do not suffer angular momentum loss in tidal effects, avoiding the catastrophe (Sommer-Larsen et al. 2003; Robertson et al. 2004; Okamoto et al. 2005; Scannapieco et al. 2008; Zavala et al. 2008; Kereš et al. 2009). For the creation of disk galaxies, it is primarily local feedback in the disk due to SNe that is thought to be most important (Sommer-Larsen et al. 1999; D'Onghia et al. 2006), though pre-heating by a UV background that mimics the reionization of the Universe also prevents gas from cooling in the lowest mass halos (Efstathiou 1992; Quinn et al. 1996; Thoul & Weinberg 1996; Navarro & Steinmetz 1997; Gnedin 2000; Hoeft et al. 2006; Okamoto et al. 2008a).

Feedback plays an additional role in the creation of galaxy disks. The heating of the gas gas by feedback delays SF and keeps gas available until low redshifts. Without feedback, halos of all masses are equally efficient at turning their cold gas into stars (Brooks et al. 2007). This leads to and early burst of SF, turning the SPH particles into collisionless star particles. This quickly uses up the gas, producing galaxies that are much too gas poor by the present day than are actually observed. Additionally, as these gas poor galaxies merge together within CDM, the predominantly collisionless particles in them relax into a spheroidal distribution, creating disk galaxies with a dominant stellar spheroid, or ellipticals. The creation of large, thin disk galaxies at low redshifts requires that gas remains available for SF until that time.

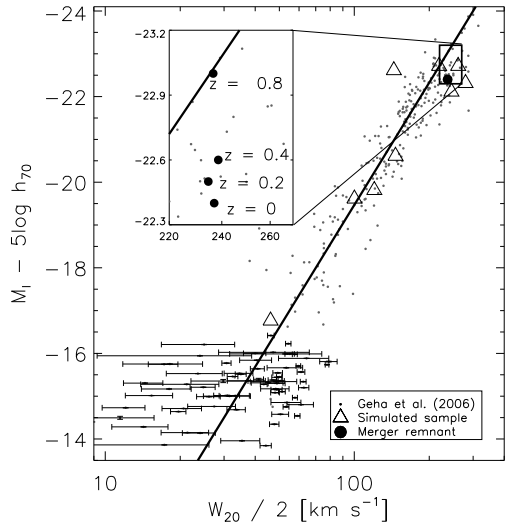

Figure 4.    The Tully-Fisher relationship, from Governato et al. (2009) for simulated disk galaxies (triangles and solid point) compared to observational data taken from Geha et al. (2006). The inset shows the evolution of one simulated galaxy as a function of redshift.

Thus, feedback is essential to 1) prevent gas from turning into stars at high z and 2) avoid artificial angular momentum loss to form large disks in the local Universe. Fig. 2 from Ceverino & Klypin (2009), demonstrates the effect that the chosen feedback prescription can have on the resulting disk. This plot again shows the circular velocity as a function of radius (i.e., the amount of mass interior to a given radius). The top panel shows the resulting distribution for a Milky Way mass progenitor that has an inefficient feedback model, resulting in the classic overcooling problem and a galaxy that is already too centrally concentrated by redshift 5. The bottom panel shows results for the same galaxy, but run with an efficient feedback model that maintains more hot gas that drives outflows and regulates star formation.

### 2.3.    How Feedback Depends on Resolution

It is important to note that SF in real galaxies is a process that takes place on small scales, of the order of parsecs. The typical force resolution in a Milky Way mass halo is instead a few hundred parsecs (e.g., Governato et al. 2007; Scannapieco et al. 2009; Okamoto et al. 2009) so that one must adopt a prescription that mimics SF on a global scale (the scheme is then termed "sub-grid"). Thankfully, SF on galaxy scales *does* appear to follow a global trend, the Kennicutt-Schmidt

Law (Kennicutt 1998; Martin & Kennicutt 2001), so that modelers can use this relation as a constraint.

However, as Fig. 2 demonstrates, even if modelers can determine the SF rate based on the Kennicutt-Schmidt law, this does not guarantee that the feedback scheme used for subsequent SN feedback will satisfy observed constraints such as the Tully-Fisher relation or the mass-metallicity relation. Early simulations introduced SN feedback only to find that the energy was quickly radiated away by the surrounding dense medium (particularly at early times when the gas densities are higher), effectively mimicking the case without feedback (e.g., Katz 1992; Steinmetz & Navarro 1999). Two main schemes have been adopted to overcome this problem. In the first, a multiphase model of the ISM is implemented (Hultman & Pharasyn 1999; Marri & White 2003; Springel & Hernquist 2003; Scannapieco et al. 2006; Harfst et al. 2006), preventing the hot gas particles from being artificially influenced by their cold gas nearest neighbors. In the second, cooling is turned off in the gas particles near a SN explosion in order to mimic the sub-resolution adiabatic expansion of the SN (Thacker & Couchman 2000, 2001). The simulations presented in the following sections, generated by members of the University of Washington's "N-Body Shop," use a SN feedback prescription that has been termed the "blastwave" scheme. This scheme also turns off cooling in nearest neighbor particles, but attempts to model this based on what is known about actual SNe, determining the radius of each SN remnant based on the analytic blastwave solution for a SN remnant (McKee & Ostriker 1977; Stinson et al. 2006), and cooling is only turned off for those particles within the blast radius. Because many SNe typically contribute feedback within a dense star forming region, the thermal energy from all of these SNe can combine to create a larger blast radius.

Due to the fact that most cosmological disk simulations achieve force resolutions of a few hundred parsecs, they do not resolve small scale density fluctuations where most SF is occurring, and must therefore adopt a prescription in which SF is allowed to occur averaged over the entire disk at very low gas densities, $\sim 0.1$ amu/cm$^{-3}$ (Governato et al. 2007; Scannapieco et al. 2009; Okamoto et al. 2009). As higher resolutions are achieved and individual star forming regions are closer to being resolved, it is important to adopt a higher density SF threshold, $\sim 100$ amu/cm$^{-3}$, that mimics the physics of actual star forming regions (Robertson & Kravtsov 2008; Tasker & Bryan 2008; Saitoh et al. 2008).

Governato et al. (2010), with a force resolution of 85pc, adopt this higher SF threshold. The high density threshold restricts SF to the small scale density fluctuations that mimic giant star forming complexes, making the SF process much "clumpier." To match the Kennicutt-Schmidt law, more star particles are born within these smaller, denser clumps. Thus, as these these stars go SN the typical amount of energy deposited into the local ISM is higher. This causes some gas within the local ISM of these clumpy regions to become unbound and escape the galaxy halo. With this prescription, significant amounts of gas are lost in feedback induced outflows, whereas this outflow is not achieved in the low SF threshold case (see also Ceverino & Klypin 2009). Fig. 3 shows $V_c$ as a function of radius for the low and high threshold cases, and the dramatic effect that SF plays in removing mass from the central portions of this galaxy.

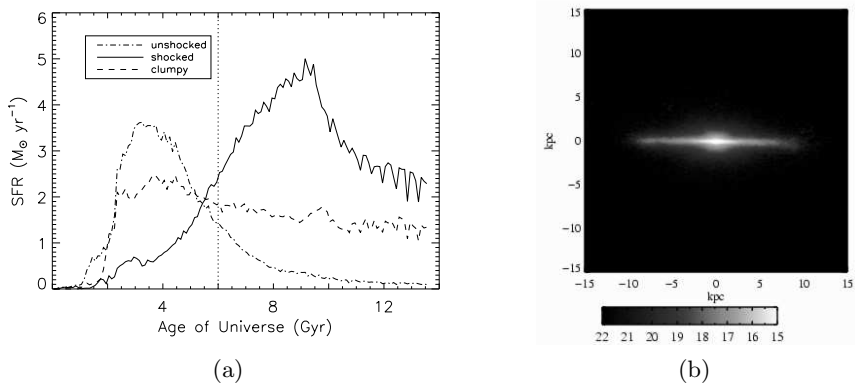

Figure 5.      From Brooks et al. (2009). *Left*: The disk SF history for a $3\times10^{12}$ M$_\odot$ halo, divided according to the accretion history of the parent gas particle. The solid line shows stars formed from gas that had been shock heated when accreted, the dot-dashed line shows stars formed from "cold flow" gas, and the dashed line shows stars formed form gas accreted in mergers. The "cold flow" gas dominates the early growth of the disk. z = 1 is marked by the vertical dotted line. *Right*: Surface brightness image (in M$_B$/arcsec$^2$) of the same galaxy, in the rest frame *B* band at z = 1. The extended disk already in place is due primarily to SF from "cold flow" gas. Image generated using SUNRISE (Jonsson 2006).

Thus, it is clear that significant progress has been made in recent years in simulating realistic disk galaxies due to the advances in resolution and feedback. The *disk* particles in recent simulations are now capable of maintaining most of their angular momentum, resulting in simulated disks that match the observed Tully-Fisher relationship, as Fig. 4 from Governato et al. (2009) shows (see also Piontek & Steinmetz 2009). This success allows us to then investigate the role that gas accretion plays in the growth of the stellar disk.

## 3.   Gas Accretion: The Impact of "Cold Flows" on Galaxy Disk Formation

There is mounting observational evidence that disk galaxies are already in place by redshift 1 (e.g., Vogt et al. 1996; Förster Schreiber et al. 2006; Stark et al. 2008; Wright et al. 2009). Studies that have attempted to investigate the size evolution of these disks have generally concluded that there has been little to no change in the size function of disk galaxies since then (Roche et al. 1998; Lilly et al. 1998; Simard et al. 1999; Ravindranath et al. 2004; Barden et al. 2005; Sargent et al. 2007; Melbourne et al. 2007; Kanwar et al. 2008). The existence of large, extended disks by z = 1 is at odds with the simplest analytic model of disk galaxy growth. Mo et al. (1998), Mao et al. (1998), and van den Bosch (1998) concluded that large disks should be unstable at high redshifts, so that large disks form since z = 1, with a factor of two increase in their size to z = 0, contrary to the observational conclusions. However, Brooks et al.

(2009) demonstrated that filamentary accretion of "cold flow" gas can lead to the growth of extended stellar disks by redshift 1, as shown in Fig. 5.

Cosmological simulations show that, contrary to the standard model in which all gas is initially shock heated to the virial temperature of a dark matter halo as it is accreted, a significant amount of gas can be accreted in galaxies without reaching the virial temperature (Fardal et al. 2001; Kereš et al. 2005, 2009). This gas has been nicknamed "cold flow" gas, although it has temperatures above a few $10^5$ K. It has been shown that the ability of a halo to shock heat gas and develop a hot gaseous halo is a function of halo mass (Forcada-Miro & White 1997; Birnboim & Dekel 2003). This trend with mass has been incorporated into most semi-analytic models (SAMs) of galaxy formation over the last couple of decades (e.g., Rees & Ostriker 1977; White & Frenk 1991; Kauffmann et al. 1993; Cole et al. 2000; Somerville et al. 2001). In these models, the cooling time for galaxy halos with masses below a few $10^{11}$ $M_\odot$ is shorter than the free fall time. Thus, the gas cools to the disk on the free fall time. Hence, our picture of how gas cools onto *low mass* galaxy disks is not altered by "cold flow" gas accretion. However, the important change to galaxy formation theory is the fact that dense filaments of gas are able to penetrate the hot halo gas of more massive galaxies, and deliver "cold flow" gas well within the virial radius of the galaxy (Dekel & Birnboim 2006; Ocvirk et al. 2008; Agertz et al. 2009; Brooks et al. 2009).

These filaments are the natural consequence of primordial density fluctuations in the early Universe, and galaxies tend to naturally form at the intersection of these filaments in the cosmic web. This filamentary gas has a shorter cooling time and can cool to the disk and form stars more rapidly than gas that was shock heated to the virial temperature. Using high resolution simulations of galaxy disks, Brooks et al. (2009) showed that, although galaxies of a Milky Way mass are dominated by *shocked* gas accretion as in the classic model, the smaller amount of "cold flow" gas in filaments dominates the formation of the stellar disk due to its short cooling time. Hence, "cold flow" gas accretion alters the standard model of disk formation for galaxies above a few $10^{11}$ $M_\odot$ in halo mass. Gas is able to cool to the disk and form stars at a higher redshift than predicted by the SAMs that do not include filamentary gas accretion.

Additionally, early SF via this "cold flow" gas can reconcile the observed evolution of the size-luminosity relationship for disk galaxies with theoretical results (Brooks et al. 2010). "Cold flows" also produce massive galaxies at higher redshift than SAMs predict. This may alleviate the discrepancy wherein massive galaxies at high z have more stars than suggested by current SAMs (Marchesini et al. 2009).

## 4. Gas Maintenance: Regulating Star Formation Efficiency with Feedback

### 4.1. Reproducing the Observed Mass-Metallicity Relation

Observationally, low mass galaxies have a lower metal abundance than their high mass counterparts. This mass-metallicity relation (MZR) is well established both locally (Tremonti et al. 2004) and at high z (Savaglio et al. 2005; Erb et al. 2006; Maiolino et al. 2008). While it has been suggested that this relationship might be due to preferential loss of metals from low mass galaxies due to SN feedback, Brooks et al. (2007) demonstrated that the gas lost from low mass galaxies was not preferentially enriched, so that adding it back into galaxies had no effect in changing the resulting MZR. Rather, low mass galaxies have lower metal abundances because they are inefficient at transforming their gas into stars that would subsequently enrich the interstellar medium.

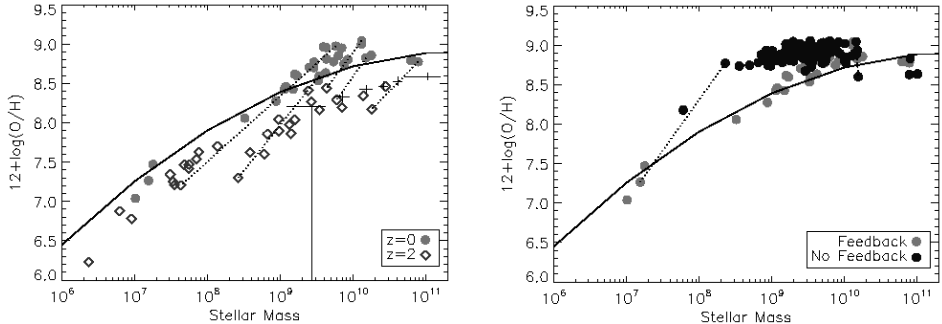

Figure 6.    The stellar mass - metallicity relationship for galaxies, from Brooks et al. (2007). *Left*: Grey circles show the results for simulated galaxies at $z=0$, while the black line shows the best fit from the observational results of Tremonti et al. (2004). Open diamonds show the simulated results at $z=2$, and the error bars are the observational results at $z=2$ from Erb et al. (2006). *Right*: Grey circles as in the left plot, but black circles show results from the same run if SN feedback is turned off. In both panels, dotted lines connect the evolution of an individual galaxy.

The left panel of Fig. 6 shows results for the stellar mass versus gas phase metallicity relationship for simulated galaxies from Brooks et al. (2007) compared to observational results. It can be seen that the simulated galaxies match the data well. This result is dependent on the nature of the SN feedback mechanism adopted. The effect of incorporating feedback is to regulate the SF as a function of mass. For low mass galaxies with their shallow potential wells, the feedback has a strong effect. It puffs up the gas, lowering the surface densities and lowering the SF efficiencies. At the high mass end, the deep potential wells allow gas effected by feedback to cool more rapidly, keeping the surface densities higher and the SF efficiencies higher. Thus, the high mass galaxies have higher SF rates, and enrich their gas to higher metallicities than the low mass galaxies, resulting in the observed trend. The right panel of Fig. 6 shows the result of turning off the SN feedback for these same galaxies. In this case, halos suffer from the overcooling problem, so that galaxies of all masses are equally efficient at making stars. Their gas is allowed to cool rapidly, forming stars quickly, and enriching the interstellar gas to the same metallicities at all halo masses, and resulting in a flat MZR.

This regulation of SF as a function of halo mass causes most of the baryons in the low mass galaxies to remain in the form of gas. Although they may lose significant amounts of baryons due to reionization or SN blowout, the gas that *remains* is inefficient at forming stars. Hence, the low mass galaxies are the most gas rich objects in the simulations. This is in agreement with observational trends (e.g., Geha et al. 2006).

Maiolino et al. (2008) demonstrated that matching the MZR has a strong dependence on the feedback prescription of the model. Models that suffer from overcooling lead to an early enrichment, and will lie above the observed MZR. The blastwave prescription continues to match the gas content of observational results at least back to $z$=3. Similar results are found at $z$=3 from Pontzen et al. (2008, 2009), who demonstrate that the blastwave mechanism can also reproduce the observed incidence rate and column density distribution of Damped Lyman Alpha absorbers in both QSO and GRB afterglow spectra. Thus, the blastwave model satisfies multiple observational constraints, and is able to reproduce the spatial and density distribution of gas in galaxies as a function of time. This result provides confidence in using this model to understand the physical processes that shape galaxy evolution.

### 4.2. Creation of a Gas Reservoir, and the Reforming of Disks

The regulation of SF by SN feedback causes the SF rates in simulated galaxies to always be less than the gas accretion rates onto the halo (Scannapieco et al. 2006; Brooks et al. 2009; Governato et al. 2009). This is a crucial result that differentiates the blastwave feedback mechanism from other models (e.g., Dekel et al. 2009a,b). It has yet to be seen what types of galaxies result at $z$=0 from the models which instead find SF rates must match gas accretion rates.

The fact that the SF rate is less than the gas accretion rate, in combination with the "cold flow" gas accretion that is allowed to rapidly cool to the disk, means that a cold gas reservoir can be built in the simulated galaxies. It has been shown that large gas fractions can lead to the rapid reformation of a disk galaxy in major merging events (Hopkins et al. 2009; Robertson et al. 2006; Robertson & Bullock 2008). This result is contrary to the long held belief that mergers either destroy or significantly thicken existing disks, making the very existence of thin disks in the local Universe a challenge within the CDM paradigm. Governato et al. (2009) show that, when cosmological gas accretion is also included, the gas fractions of the progenitor galaxies need not be more than 25%. For a merger of 2:1 mass ratio at $z$=1, 1/3 of the newly formed stellar disk mass at $z$=0 forms from gas accumulated in the cold gas reservoir prior to the last major merger, but 2/3 is formed from gas that is accreted after the merger or able to cool from the hot gas halo. Thus, cold flows play a significant role in contributing to the cold gas reservoir that can reform a disk, but subsequent gas accretion onto the halo cannot be neglected.

### 5. Gas Loss: The Origin of Bulgeless Galaxies

Roughly 50% of star forming galaxies at $10^{10}$ $M_\odot$ in stellar mass have no central bulge component, and the number of bulgeless galaxies rises further to lower masses (Dutton 2009). Thus, the centrally concentrated mass profiles in simulations that were discussed in Section 2 have remained a particular challenge to CDM theory, preventing the simulation of a bulgeless galaxy. However, as Fig. 3 demonstrated, the adoption of a physically motivated SF and SN feedback prescription has a larger role in decreasing the central mass concentrations than resolution. Hence, the inability to simulate a bulgeless disk galaxy appears to be due to missing physics.

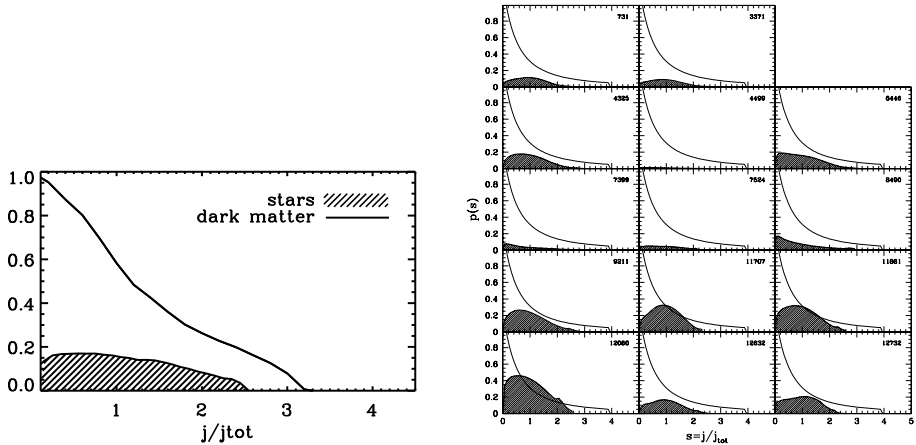

Figure 7.    *Left*: The angular momentum distribution of disk stars and dark matter for a simulated bulgeless disk galaxy, from Governato et al. (2010). *Right*: The observationally derived angular momentum distribution for low mass local disk galaxies, from van den Bosch et al. (2001), reproduced by permission of the authors.

In particular, theoretical work has shown that the physics required to create a bulgeless disk galaxy must be able to modify the distribution of angular momentum of star forming gas. The angular momentum per unit mass, $j$, of gas can be conserved during halo collapse so that it follows that of the dark matter (van den Bosch et al. 2002; D'Onghia et al. 2006; Zavala et al. 2008). While this appears to be a necessary condition, it is not sufficient to reproduce observations. If the gas maintains this distribution, the overall amount of low $j$ gas will be too large, leading to bulges much larger than observed in the real Universe (van den Bosch et al. 2001; Bullock et al. 2001; Maller & Dekel 2002; van den Bosch et al. 2002, 2003; D'Onghia & Burkert 2004). Previous simulations aimed to *maintain* the angular momentum of the gaseous component, and in doing so consistently produced galaxies with large bulges.

Bulgeless disks are more common in low mass dwarf galaxies (Dutton 2009), where supernova feedback plays a stronger role due to their shallower potential wells. This argues that feedback may be the main culprit in altering the angular momentum distribution of the baryons after halo collapse. In low mass galaxies with their shallow potential wells, as low $j$ material settles to the center of the galaxy and undergoes SF, subsequent feedback can remove the surrounding low $j$ gas completely from the galaxy in outflows, so that the angular momentum distribution no longer follows that of the dark matter. At the high mass end, deeper potential wells prevent the complete loss of low $j$ material, so that these galaxies will still form bulges (though some low $j$ material may be lost at early times, reducing the overall bulge mass). This picture allows for the trend in bulge prominence as a function of galaxy mass, yielding a solution to the mismatch between theory and observations.

As seen in Section 2.3, simulators continue to push to higher resolutions as computational power continues to increase, allowing for a more realistic treatment of SF and SN feedback. In the case of the low mass disk galaxy in Gov-ernato et al. (2010), high resolution and realistic SF allow for gas outflows that significantly reduce the central mass concentration of this simulated galaxy. The outflowing gas has much lower $j$ than the baryons that remain in the galaxy. This low $j$ gas is the gas that lies at the center of the galaxy and, if retained, would form a large stellar bulge. However, the loss of this gas leads instead to a stellar disk with a purely exponential surface brightness profile. This disk would be classified as "bulgeless" observationally.

Fig. 7 compares the resulting angular momentum distribution of the stars and dark matter in Governato et al. (2010) (left panel) to the observationally derived probability distributions for local disk galaxies (right panel, from van den Bosch et al. 2001). It can be seen that the simulated bulgeless disk galaxy has a distribution in excellent agreement with the observed low mass disk galaxies. The black lines in the observational results show the expected distribution if the baryons follow that of the dark matter. The main mismatch between theory and observations lies at both the low $j$ and high $j$ ends of the distribution. For the simulated galaxy, the low $j$ material has been lost in feedback induced outflows, while the high $j$ gaseous material lies outside the star forming region of the disk.

Finally, the loss of gas in outflows has a significant effect on the dark matter distribution of the simulated galaxy. Mashchenko et al. (2006) and Mashchenko et al. (2008) showed that bulk gas motion created by highly resolved star formation and feedback can alter the central potential of a galaxy on a timescale comparable to the crossing time of the dark matter particles. In this case, the kinetic energy of the dark matter is altered. When the central concentration of baryons is removed in outflows, the binding energy is decreased, allowing the dark matter to relax to a new configuration and flattening the density profile (Navarro et al. 1996; El-Zant et al. 2001; Mo & Mao 2004; Read & Gilmore 2005; Tonini et al. 2006; Dutton et al. 2007).

## 6. Conclusions

In summary, we have seen that achieving a successful model for the evolution of gas in galaxies leads to simulated galaxies that more closely match observations. In particular, the models presented here can build large disks at early times due to filamentary gas accretion, match the Tully-Fisher relation, match the stellar mass - metallicity relation as a function of time, reproduce the observed trends in gas fraction with mass and time, produce disks despite low z major mergers, create gas outflows that lead to bulgeless disk galaxies and an angular momentum distribution similar to that observed.

The success of these simulations is due largely to their feedback mechanism. After the gas is accreted, it is imperative in these models that subsequent SF be regulated. The successful models currently rely heavily on SN feedback that regulates SF efficiency as a function of mass.

Of course, many avenues remain unexplored, and there are still challenges to the simulators. It has yet to be seen if the simulations can reproduce the luminosity function of galaxies as a function of redshift. It has also been noted that current simulations continue to form too many stars (Guo et al. 2009). This situation may be remedied in the future by inclusion of a more realistic model for SF that includes molecular cooling (rather than the atomic cooling used by many simulators) and the formation of stars from molecular gas. Simultaneously reproducing the Tully-Fisher relation and the luminosity function is thus a serious and necessary constraint. Furthermore, it has yet to be demonstrated that the model that successfully produces a bulgeless disk in dwarf galaxies will also produce bulges at the massive end, as required by observations. Simulating a Milky Way mass galaxy at the desired resolution ($<100$pc) requires a larger simulation volume and significantly more particles, making this currently a computational challenge. Finally, the results presented here focus on Milky Way mass galaxies or smaller, but a coherent model must be developed that extends to higher masses, and the formation of ellipticals and clusters. This will require modeling of black hole growth and AGN feedback in a cosmological setting (e.g., Okamoto et al. 2008b), a field just beginning to emerge.

**Acknowledgments.**    AB would like to thank Fabio Governato, Justin Read, and Simon White for their useful comments on these proceedings.